\documentstyle[preprint,aps]{revtex}

\begin{document}
\draft
\author{D. Galetti}
\address{Instituto de F\'{\i}sica{}{} Te\'{o}rica, Universidade Estadual Paulista -\\
UNESP \ \\
Rua Pamplona 145 \\
01405 - 900 S\~{a}o Paulo, S.P.; Brazil}
\author{A.F.R. de Toledo Piza}
\address{Instituto de F\'{\i}sica, Universidade de S\~{a}o Paulo \\
C.P. 66318, 05315-970, S\~{a}o Paulo, S.P.; Brazil}
\title{Giant monopole resonance and nuclear compression modulus for $^{40}$Ca and $%
^{16}$O}
\maketitle

\begin{abstract}
Using a collective potential derived on the basis of the Generator
Coordinate Method with Skyrme interactions we obtain values for the
compression modulus of $^{40}$Ca which are in good agreement with a recently
obtained experimental value. Calculated values for the compression modulus
for $^{16}$O are also given. The procedure involved in the derivation of the
collective potential is briefly reviewed and discussed.
\end{abstract}


\section{Introduction.}

Among the collective vibrational modes of nuclei, the isoscalar giant
monopole resonance (GMR), also called breathing mode, has received
considerable attention due to the fact that its energy is related to the
nuclear compression modulus. This property corresponds, in the limit of an
infinite number of nucleons, to the nuclear matter incompressibility, which
is of great interest in areas such as heavy ion collisions and nuclear
astrophysics. The states corresponding to this type of collective motion
have long been experimentally identified \cite{young1} and various
microscopic theoretical approaches have been developed in order to account
for and improve the deduced values of the compression modulus \cite
{blaizot1,blaizot2,flovau,treiner}. In recent years, more accurate
experimental determination of the GMR energies \cite{young2} led to renewed
interest in the determination of the nuclear matter incompressibility.
Widely ranged theoretical approaches have been employed to obtain the
compression modulus of finite nuclei and of nuclear matter, including
microscopic calculations with the Gogny interaction \cite{blaizot3} and with
generalized Skyrme forces \cite{farine}, self-consistent Hartree-Fock plus
Random Phase Approximation treatments with Skyrme interactions \cite
{hamamoto}, relativistic mean field calculations \cite{chossy}, a
Thomas-Fermi \cite{myers} and a Fermi liquid drop calculations \cite{kolo}.

Motivated by the recent experimental results \cite{young2}, and by the
discussion of the compression modulus in finite nuclei presented in \cite
{young}, we report on a microscopic calculation of the compression modulus
for $^{40}$Ca and $^{16}$O based on the Generator Coordinate Method (GCM) 
\cite{hillwhee,griwhee,wong}. This method allows for the microscopic
determination of a collective potential $V(q)$, $q$ being the appropriate
collective coordinate for the breathing mode, from which the compression
modulus can be calculated directly as

\[
K=\left. \frac{1}{A}\frac{d^{2}}{dq^{2}}V\left( q\right) \right| _{q=q_{0}}, 
\]
where $A$ is the nuclear mass number and $q_{0}$ corresponds to the minimum
of $V(q)$. This approach has been developed in its general aspects many
years ago \cite{dipi} and can be directly used in the present context. We
show in continuation that the collective potential does not coincide with
the diagonal part of the GCM energy kernel due to non-negligible
off-diagonal contributions; moreover, the correlations embodied in the
choice of the generator coordinate introduce anharmonicities in the
collective potential, with the result that the deduced compression modulus
comes out higher than that obtained from the standard variational approach.

In our calculations we have used Skyrme interactions without Coulomb forces.
Values for the compression modulus were obtained for $^{40}$Ca and for $%
^{16} $O. In the case of the heavier nucleus there is good agreement with
the available experimental data especially when the Skyrme interactions SII
and SIII are used. The GCM results also give, in the case of the lighter
nucleus, values for the compression modulus which are only moderately lower
than the values obtained for the heavier nucleus. Skyrme forces without
density dependence such as SV lead to values for the compression modulus
which are appreciably lower in both cases.

In section II the GCM and its connection with collective Hamiltonians is
briefly reviewed. In section III the collective potential for the breathing
mode and the microscopically deduced expression for the compression modulus
are derived. Values of the incompressibility are calculated for five sets of
Skyrme parameters. Finally, in section IV we present our conclusions.

\section{GCM kernels and collective Hamiltonians.}

The well known Generator Coordinate Method is based on an ansatz for the
collective wave function which is set up as

\[
|\Psi \left( \tilde{r}\right) \rangle =\int f\left( \alpha \right) \;\Phi
\left( \tilde{r},\alpha \right) \;d\alpha =\int f\left( \alpha \right)
\;|\alpha \rangle \;d\alpha , 
\]
where the $\{\Phi\left(\tilde{r},\alpha\right)\}$ (or, for short, the $%
\{|\alpha \rangle\}$) constitute a set of auxiliary nuclear many-body wave
functions parametrized by $\alpha $ (the generator coordinate), related to
the nuclear degree of freedom one wishes to describe, $f\left(\alpha\right) $
is a weight function to be determined variationally and $\tilde{r}$ is a
shorthand notation for all the nucleon spatial coordinates. One way of
constructing the $\Phi\left(\tilde{r},\alpha \right) $ is through the use of
Slater determinants of the lowest single particle eigenfunctions $%
\varphi\left(\vec{r};\alpha\right)$ of an ad hoc $\alpha$-dependent
auxiliary potential. The choice of the $\alpha$-dependence of the auxiliary
potential should therefore suit the particular collective degree of freedom
under consideration (e.g. an overall scale parameter in the case of the
breathing mode).

For a given nucleon-nucleon interaction, we can write the nuclear many-body
Hamiltonian $H$ and use the Rayleigh-Ritz variational principle to obtain
the so-called Griffin-Wheeler equation\cite{griwhee} for the weight funtion $%
f(\alpha )$

\begin{equation}
\int \left[ \langle \alpha |H|\alpha ^{\prime }\rangle -E\langle \alpha
|\alpha ^{\prime }\rangle \right] f\left( \alpha ^{\prime }\right) d\alpha
^{\prime }=0,  \label{gri}
\end{equation}
where $E$ is introduced as a Lagrange multiplier to account for wave
function normalization. In this equation the energy kernel $\langle \alpha
|H|\alpha ^{\prime }\rangle $ and the overlap kernel $\langle \alpha |\alpha
^{\prime }\rangle $ are functions of $\alpha $ and $\alpha ^{\prime }$ only,
the nucleon coordinates having been integrated out.

In principle, the solutions of the integral eigenvalue equation (\ref{gri})
give a variational ground state and a set of collective excited states
expressed in terms of weight functions $f_{i}(\alpha )$ and the
corresponding eigenvalues $E_{i}$. Technical problems related to the overlap
properties of the auxiliary many-body functions $|\alpha \rangle $ in many
cases prevent the implementation of such a direct approach, however. A
discussion of these difficulties and of general methods to circunvent them
can be found in \cite{pieme,ring}, where it is shown that by applying
general transformations to the Griffin-Wheeler equation the determination of
the collective spectrum can be reduced to a standard eigenvalue problem.

A convenient choice for the nucleon-nucleon interaction is the Skyrme force 
\cite{skyrme}, which contains two body and three body terms. The two-body
term is

\begin{eqnarray*}
v_{1,2}\left(\vec{r}_{1},\vec{r}_{2}\right) &=&t_{0}\left( 1+x_{0}P_{\sigma
}\right) \delta \left(\vec{r}_{1}- \vec{r}_{2}\right)  \nonumber \\
&&+\frac{t_{1}}{2}\left[ \delta \left( \vec{r}_{1}-\vec{r}_{2}\right) \vec{k}%
^{2}+ \vec{k^{\prime }}^{2}\delta \left( \vec{r}_{1}- \vec{r}_{2}\right) %
\right]  \nonumber \\
&&+ t_{2}\vec{k^{\prime }}\cdot \delta \left( \vec{r}_{1}- \vec{r}%
_{2}\right) \vec{k}  \nonumber \\
&& +iW_{0}\left( \vec{\sigma }_{1}+\vec{\sigma _{2}}\right) \cdot \vec{%
k^{\prime }}\times \delta \left( \vec{r}_{1}- \vec{r}_{2}\right) \vec{k},
\end{eqnarray*}
where $t_{0}$, $x_{0}$, $t_{1}$, $t_{2}$ and $W_{0}$ are adjusted
parameters, and $P_{\sigma }$ is the spin exchange operator. The three-body
part of the Skyrme force is simply taken as a zero range Wigner force

\[
v_{1,2,3}\left( \vec{r}_{1},\vec{r}_{2},\vec{r}_{3}\right) =t_{3}\delta
\left( \vec{r}_{1}-\vec{r}_{2}\right) \delta \left( \vec{r}_{2}-\vec{r}%
_{3}\right) , 
\]
where $t_{3}$ is an additional parameter. For even nuclei this contribution
can be rewritten as the density dependent two-body interaction

\[
v_{1,2,3}\left( \vec{r}_{1},\vec{r}_{2}\right) =\frac{ t_{3}}{6}\left(
1+P_{\sigma }\right) \delta \left( \vec{r}_{1}- \vec{r}_{2}\right) \rho
\left( \frac{\vec{r}_{1}+ \vec{r}_{2}}{2}\right). 
\]

Since we are interested in the monopole giant resonance spectrum of
spherical nuclei, we choose an oscillator potential with the oscillator
parameter $\beta$ as generator coordinate, which is thus directly related to
the nuclear radius. The auxiliary many-body functions $|\beta\rangle$ will
therefore be Slater determinants built from single particle oscillator
states $\left\{ \varphi_\lambda \left( \vec{r};\beta \right) \right\}$ with
oscillator parameter $\beta $. The corresponding overlap kernel then reads 
\cite{gri,flovau}

\[
\langle \beta |\beta ^{\prime }\rangle =\left( \frac{2\beta \beta ^{\prime } 
}{\beta +\beta ^{\prime }}\right) ^{T}, 
\]
where $T$ is $6$, $36$ and $120$ for $^{4}$He, $^{16}$O and $^{40}$Ca
respectively. In order to handle this overlap kernel it is convenient to
perform a change of variable by writing $\beta =\beta _{0}\exp \left( \alpha
\right)$, where $\beta _{0}$ defines a reference length scale and $\alpha $
is a new generator coordinate. With this choice the overlap kernel appears as

\[
\langle \alpha |\alpha ^{\prime }\rangle ={\rm sech}^{T}\left( \alpha
-\alpha ^{\prime }\right) 
\]
which, for the heavier nuclei, can be approximated by the Gaussian

\begin{equation}
\langle \alpha |\alpha ^{\prime }\rangle \simeq \exp \left[ -\frac{T}{2}%
\left( \alpha -\alpha ^{\prime }\right) ^{2}\right] .  \label{over}
\end{equation}

The energy kernel can be put in the form \cite{flovau} 
\begin{eqnarray}
\langle \alpha |H|\alpha ^{\prime }\rangle &=&\langle \alpha |\alpha
^{\prime }\rangle \times \left[ C_{1}{\rm sech}\left( \alpha -\alpha
^{\prime }\right) \exp \left( \alpha +\alpha ^{\prime }\right) \right] 
\nonumber \\
&&+C_{2}\cosh ^{3/2}\left( \alpha -\alpha ^{\prime }\right) \exp \left[ 
\frac{3}{2}\left( \alpha +\alpha ^{\prime }\right) \right]  \nonumber \\
&&+C_{3}\cosh ^{5/2}\left( \alpha -\alpha ^{\prime }\right) \exp \left[ 
\frac{5}{2}\left( \alpha +\alpha ^{\prime }\right) \right]  \nonumber \\
&&+C_{4}\cosh ^{3}\left( \alpha -\alpha ^{\prime }\right) \exp \left[
3\left( \alpha +\alpha ^{\prime }\right) \right] ,  \label{eker}
\end{eqnarray}
where the coefficients $C_{i},\;\,i=1$ to $4$ are given by

\[
C_{1}=\frac{\hbar ^{2}}{2m}a_{1},\hspace{1cm} C_{2}=\frac{3}{2}\frac{t_{0}}{%
\left( 2\pi \right) ^{3/2}}a_{2}, 
\]

\[
C_{3}=\frac{3t_{1}+5t_{2}}{8\left( 2\pi \right) ^{3/2}}a_{3}+\frac{
9t_{1}-5t_{2}}{16\left( 2\pi \right) ^{3/2}}a_{4},\hspace{1cm} C_{4}=4\frac{%
t_{3}}{\left( \pi \sqrt{3}\right) ^{3}}a_{5}, 
\]
and $a_{1}$, $a_{2}$, $a_{3}$, $a_{4}$, and $a_{5}$ are constants resulting
from the integrations over the densities, having therefore values specific
to each nucleus.

Attempts to solve directly the resulting Griffin-Wheeler equation run into
instability problems in the determination of the weight functions $f\left(
\alpha \right)$. They can however be circumvented by means of an analytical
procedure discussed by Piza and Passos\cite{pieme}. It consists in
transforming the Griffin-Wheeler equation by introducing the new set of
states

\[
|k\rangle =\int \frac{U_{k}\left( \alpha \right) }{\Lambda ^{1/2}\left(
k\right) }|\alpha \rangle d\alpha, 
\]
where $U_{k}\left( \alpha \right) $ and $\Lambda \left( k\right) $ are
defined through the diagonalization process for the overlap kernel

\begin{equation}
\int U_{k}^{\dagger }\left( \alpha \right) \langle \alpha |\alpha ^{\prime
}\rangle U_{k^{\prime }}\left( \alpha ^{\prime }\right) d\alpha d\alpha
^{\prime }=\Lambda \left( k\right) \delta \left( k-k^{\prime }\right) .
\label{diag}
\end{equation}
Since the states $|k\rangle $ are orthonormal and complete\cite{pieme} we
are led to the new equation

\[
\int \left[ H\left( k,k^{\prime }\right) -E\right] g\left( k^{\prime
}\right) dk^{\prime }=0, 
\]
where the new energy kernel is written as

\begin{equation}
H\left( k,k^{\prime }\right) =\langle k|H|k^{\prime }\rangle =\int \frac{%
U_{k}^{\dagger }\left( \alpha \right) }{\Lambda ^{1/2}\left( k\right) }%
\langle \alpha |H|\alpha ^{\prime }\rangle \frac{U_{k^{\prime }}\left(
\alpha ^{\prime }\right) }{\Lambda ^{1/2}\left( k^{\prime }\right) }d\alpha
d\alpha ^{\prime },  \label{nhk}
\end{equation}
and

\[
g\left( k\right) =\Lambda ^{1/2}\left( k\right) \int U_{k}^{\dagger }\left(
\alpha \right) f\left( \alpha \right) d\alpha . 
\]
For translationally invariant kernels (i.e., depending on $\alpha -\alpha
^{\prime }$ only) such as (\ref{over}), the diagonalization (\ref{diag}) is
performed simply by a Fourier transform, so that $\Lambda \left( k\right) $
is also a Gaussian, and there are no longer harmfull divergences. The
transformations can be directly performed on Eq.(\ref{nhk}) giving the new
energy kernel in terms of the Fourier variable $k$.

At this stage, numerical calculations can be performed and the energies and
wave functions of the breathing mode can be obtained \cite{flovau,dipi}. The
resulting collective spectrum, obtained using the Skyrme interaction without
the Coulomb interaction, is known \cite{flovau,dipi} and needs no further
discussion.

\section{Collective potential and compression modulus.}

The point we want to stress in this note is that a value for the nuclear
compression modulus can be deduced from the energy kernel given by Eq.(\ref
{nhk}). Although at first sight one is tempted to associate a collective
potential to the diagonal part of the GCM energy kernel $\langle \alpha
|H|\alpha ^{\prime }\rangle $, we will show that off-diagonal elements also
contribute significantly to the properly defined collective potential. Since
the energy kernel has the overlap kernel $\langle \alpha |\alpha ^{\prime
}\rangle $ as a global factor (see Eq.(\ref{eker})) and the overlap kernel
is narrower for heavier nuclei, the contributions due to off-diagonal
elements will be relatively more significant for lighter nuclei.

In order to extract a collective potential from the transformed energy
kernel $H\left( k,k^{\prime }\right) $, Eq. (\ref{nhk}), we follow a method
presented many years ago \cite{dipi}. The first step consists of performing
a double Fourier transform on $H\left( k,k^{\prime }\right) $

\[
H\left( k,k^{\prime }\right) 
\mathrel{\mathop{\rightarrow }\limits_{{\cal F}}}%
H\left( x,x^{\prime }\right). 
\]
The resulting non-local energy function is then subjected to a Weyl-Wigner
transformation

\[
h\left( q,p\right) =\int \langle q-\frac{\sigma }{2}|H|q+\frac{\sigma }{2}
\rangle \exp \left( ip\frac{\sigma }{\hbar }\right) d\sigma , 
\]
where we have introduced the new variables

\[
q = \frac{x+x^{\prime}}{2},\hspace{1cm}\sigma=x^{\prime}-x. 
\]
It is then clear that the nonlocality of $H\left( x,x^{\prime }\right)$
gives rise to the momentum dependence of the collective Hamiltonian. If we
expand in the nonlocality parameter, $\sigma $, we obtain the series

\[
H\left( q,\sigma \right) =\langle q-\frac{\sigma }{2}|H|q+\frac{\sigma }{2}%
\rangle =\sum_{n=0}^{\infty }H^{\left( n\right) }\left( q\right) \delta
^{\left( n\right) }\left( \sigma \right) , 
\]
where the $H^{(n)}$ coefficients are $n$-th moments of the energy kernel

\[
H^{\left( n\right) }\left( q\right) =\frac{\left( -1\right) ^{n}}{n!}\int
H\left( q,\sigma \right) \sigma ^{n}d\sigma. 
\]
The resulting Weyl-Wigner energy function

\[
h\left( q,p\right) =\sum_{n=0}^{\infty }H^{\left( n\right) }\int \delta
^{\left( n\right) }\left( \sigma \right) \exp \left( ip\frac{\sigma }{\hbar }
\right) d\sigma 
\]
is associated to the collective Hamiltonian operator

\[
H\left( \widehat{q},\widehat{p}\right) =H^{\left( 0\right) }\left( \widehat{%
q }\right) +\sum_{n=1}^{\infty }\frac{\left( -i\right) ^{n}}{\left( 2\hbar
\right) ^{n}}\left\{ \left\{ ...\left\{ H^{\left( n\right) }\left( \widehat{%
q }\right) ,\widehat{p}\right\} \right\} ...\right\} 
\]
in which the $p$-independent term $H^{\left( 0\right) }\left( \widehat{q}%
\right)$ is interpreted as the collective potential, namely

\[
V\left( \widehat{q}\right) \equiv H^{\left( 0\right) }\left( \widehat{q}%
\right) =\int H\left( q,\sigma ^{\prime }\right) d\sigma ^{\prime }. 
\]
We will not discuss the inertia associated to the breathing mode, since this
has been already before \cite{dipi}. In what follows, we will concentrate
instead on the collective potential.

In order to evaluate $V\left( \widehat{q}\right)$ let us return to the GCM
expressions and expand the reduced kernel $H\left( \alpha ,\alpha ^{\prime
}\right) /N\left( \alpha ,\alpha ^{\prime }\right) $ (which can be read
directly from Eq. (\ref{eker})) as follows. Introducing the variables $\eta
=\alpha -\alpha ^{\prime }$, and $\gamma =\frac{\alpha +\alpha ^{\prime }}{2}
$, the general term of the series expansion of this object around the
minimum $\gamma _{0}$ of its the diagonal part is

\[
C_{mn}\eta^m(\gamma-\gamma_0)^n=\frac{1}{n!m!}\frac{\partial ^{n+m}}{%
\partial \eta ^{m}\partial \gamma ^{n}}\left[ \frac{H\left( \alpha ,\alpha
^{\prime }\right) }{N\left( \alpha ,\alpha ^{\prime }\right) }\right]
_{\gamma =\gamma _{0},\eta =0}\eta^m(\gamma-\gamma_0)^n. 
\]
As a result of the symmetry of the GCM kernels only even values of $m$ and $%
n $ appear in this expansion. The collective potential $V(q)$ is
correspondingly also given as a sum of terms $G_{mn}(q-\gamma_0)$ which can
be expressed as

\begin{equation}
G_{mn}\left(q-\gamma_0\right) =\frac{C_{mn}}{2\pi }\int \exp \left( \frac{-T%
}{2}\eta^{2}\right)\eta^{m}d\eta\int\int \frac{\exp \left[
ik\left(\gamma-q\right) \right] }{\Lambda \left( \frac{k}{2}\right) }%
(\gamma-\gamma_0)^{n}dkd\gamma.
\end{equation}
This is a somewhat symbolic expression, the meaning of which is specified by
the prescription that the $\gamma$ integration is to be performed first in
terms of derivatives of the delta function $\delta(k)$, so that the integral
over $k$ gives derivatives of the $k$-dependent part of the remaining
integrand evaluated at $k=0$. Following this procedure one may collect the
powers of $(q-\gamma_0)$ and write the collective potential as

\[
V\left(q\right)=\sum_{\nu}\;V_{\nu}\times\left( q-\gamma _{0}\right)^{\nu}. 
\]
Numerical results indicate sufficient convergence when this sum is truncated
at $\nu=6$ in the two cases studied here. The relevant coefficients $V_{\nu}$
are given by

\begin{eqnarray}  \label{v2}
V_{0}&=&C_{00}-\frac{1}{4T}C_{02}+\frac{1}{T}C_{20}+\frac{3} {16T^{2}}C_{04}+%
\frac{3}{T^{2}}C_{40}-\frac{1}{4T^{2}}C_{22} -\frac{15}{64T^{3}}C_{06}+\frac{%
15}{T^{3}}C_{60}  \nonumber \\
&&+\frac{3}{16T^{3}}C_{24}-\frac{3}{4T^{3}}C_{42}  \nonumber \\
\nonumber \\
V_{1}&=&-\frac{3}{4T}C_{03}+\frac{1}{T}C_{21}+\frac{5}{16T^{2}} C_{05}-\frac{%
3}{4T^{2}}C_{23}+\frac{3}{T^{2}}C_{41}  \nonumber \\
\nonumber \\
V_{2}&=&C_{02}-\frac{3}{2T}C_{04}+\frac{1}{T}C_{22}+\frac{45} {16T^{2}}%
C_{06}-\frac{3}{2T^{2}}C_{24}+\frac{3}{T^{2}}C_{42} \\
\nonumber \\
V_{3}&=&C_{03}-\frac{5}{2T}C_{05}+\frac{1}{T}C_{23}  \nonumber \\
\nonumber \\
V_{4}&=&C_{04}-\frac{15}{4T}C_{06}+\frac{1}{T}C_{24}  \nonumber \\
\nonumber \\
V_{5}&=&C_{05}  \nonumber \\
\nonumber \\
V_{6}&=&C_{06}.  \nonumber
\end{eqnarray}

The above results show explicitely that the collective potential does in
fact contain contributions from the off-diagonal terms of the GCM energy
kernel, as mentioned before. In fact, the diagonal part generates only the
terms $C_{0n}$. Furthermore, it is also evident that the minimum of the
collective potential does not coincide with that of the diagonal part of the
GCM energy kernel due to the off-diagonal contributions, which therefore
introduce corrections to the equilibrium radius of the nucleus. Furthermore,
we obtain the main result that the compression modulus, defined by

\[
K=\left. \frac{1}{A}\frac{d^{2}}{dq^{2}}V\left( q\right) \right| _{q=q_{0}}, 
\]
where $q_{0}$ is the minimum of the colletive potential, does not coincide
with the value obtained when the standard variational principle (based on
the use of the diagonal part only of the GCM energy kernel) when the
oscillator parameter is taken as the variational parameter. In fact, we
obtain from the colletive potential

\begin{eqnarray}
K&=&\frac{2}{A}\left[ V_{2}+3V_{3}\left( q-\gamma _{0}\right) +6V_{4}\left(
q-\gamma _{0}\right) ^{2} \right.  \nonumber \\
&&+ \left. 10V_{5}\left( q-\gamma _{0}\right) ^{3}+15V_{6}\left( q-\gamma
_{0}\right) ^{4}\right] _{q=q_{0}},  \label{k}
\end{eqnarray}
whereas on the basis of the simpler variational method we would have

\begin{equation}  \label{stvar}
K_{v}=\frac{2}{A}C_{02}.
\end{equation}
Comparison with Eq.(\ref{v2}) reveals the off-diagonal contributions to the
predicted value of the compression modulus.

Numerical results for $^{40}$Ca using Eq.(\ref{k}) are shown in Table I
using five different sets of Skyrme parameters, labelled as SI to SV. The
values corresponding to the standard variational procedure (\ref{stvar}) are
also shown for comparison.

\[
\stackrel{\text{{\large Table I - Compression modulus for} }^{\text{{\large %
40}}}\text{{\large Ca}}}{
\begin{tabular}{||l|l|l|l|l|l||}
\hline\hline
Compression 
\mbox{$\backslash$}%
Interaction & SI & SII & SIII & SIV & SV \\ \hline
K (MeV) & 253.84 & 229.08 & 239.04 & 217.86 & 205.37 \\ \hline
K$_{v}$ (MeV) & 245.87 & 222.28 & 231.73 & 211.57 & 199.64 \\ \hline\hline
\end{tabular}
} 
\]
It can be seen that the compression modulus obtained from the collective
potential is higher than that calculated through the simple variational
method. In fact, it is interesting to observe that the results obtained with
interactions SII and SIII are in good agreement with the experimental value
presented by Youngblood\cite{young}.

Using the sixth-order approximation to the collective potential we can also
calculate the compression modulus for $^{16}$O. The results are shown in
Table II below.

\[
\stackrel{\text{{\large Table\ II\ -\ Compression\ modulus\ for}\ }^{\text{%
{\large 16}}}\text{\ {\large O}}}{
\begin{tabular}{||l|l|l|l|l|l||}
\hline\hline
Compression 
\mbox{$\backslash$}%
Interaction & SI & SII & SIII & SIV & SV \\ \hline
K (MeV) & 242.02 & 213.78 & 223.93 & 202.60 & 190.48 \\ \hline
K$_{v}$ (MeV) & 216.15 & 192.17 & 200.56 & 182.71 & 172.47 \\ \hline\hline
\end{tabular}
} 
\]
In this case the value of the compression modulus obtained from the
collective potential is about $10\%$ higher than that obtained from the
simple variational calculation. This indicates a theoretical value for the
compression modulus of $^{16}$O which is not much lower than the observed
experimental values for heavier nuclei \cite{young}. It is also worth
noticing that for SV, which does not include the three-body force and
therefore does not present a density dependent term, the compressibility
modulus is lower in both cases.

\section{Concluding remarks.}

Values for the compression moduli of finite nuclei have been obtained from a
collective potential associated to a description of the breathing modes
based on the Generator Coordinate Method. The procedure involved in the
derivation of the collective potential from the GCM kernels has been
reviewed. It has been shown that this procedure leads to a collective
potential that embodies contributions coming from off-diagonal elements of
the GCM kernels. They are associated to correlations introduced by the
choice of generator coordinate and by the use of the Griffin-Wheeler ansatz,
which lead to results which differ from those obtained from simpler
variational procedures. In particular, values of nuclear radii and nuclear
densities are different in both cases\cite{hodgson}.

This approach has been applied to the calculation of the nuclear
incompressibility for $^{40}$Ca and $^{16}$O, using Skyrme effective
interactions without Coulomb forces. The results are, as expected, higher
than those obtained from simple variational approaches and, for interactions
SII and SIII, are in good agreement with the experimental results given by
Youngblood et al\cite{young} for $^{40}$Ca. Our results for the lighter
nucleus $^{16}$O give values which are not too low when compared to the
available experimental values for heavier nuclei and to the mean value of $%
231\pm 5$ MeV assigned to nuclear matter by Youngblood.

\vspace{.3cm}

\noindent {\bf Acknowledgment: } D.G. is partially supported by Conselho
Nacional de Desenvolvimento Cient\'{\i}fico e Tecnol\'ogico (CNPq), Brazil.

\end{document}